\documentclass[preprint,pra,showpacs,nofootinbib]{revtex4}
\usepackage[mathscr]{euscript}
\usepackage{dcolumn}
\usepackage{bm}
\usepackage{graphicx}
\usepackage{subfigure}
\usepackage{slashbox}
\usepackage{dcolumn}
\usepackage{amsmath,amssymb,amsthm}
\usepackage[colorlinks=true,linkcolor=red]{hyperref}
\newcommand{\bea}{\begin{eqnarray}}
\newcommand{\eea}{\end{eqnarray}}
\newcommand{\beq}{\begin{equation}}
\newcommand{\eeq}{\end{equation}}

\def\/{\over}

\begin{document}

\title{CMB power spectrum for emergent scenario and slow expansion in scalar-tensor theory of gravity}
\author{Qihong Huang$^{1}$\footnote{Corresponding author: huangqihongzynu@163.com}, He Huang$^{2}$ and Bing Xu$^{3}$ }
\affiliation{
$^1$ School of Physics and Electronic Science, Zunyi Normal University, Zunyi 563006, China\\
$^2$ Institute of Applied Mechanics, Zhejiang University, Zhejiang 310058, China\\
$^3$ School of Electrical and Electronic Engineering, Anhui Science and Technology University, Bengbu, Anhui 233030, China
}

\begin{abstract}
We analyze the stability of the Einstein static universe in scalar-tensor theory of gravity, and find it can be stable against both scalar and tensor perturbations under certain conditions. By assuming the emergent scenario originating from an Einstein static state, followed by an instantaneous transition to an inflationary phase, we study and obtain the analytical approximations of the primordial power spectrum for the emergent scenario. Then, we plot the primordial power spectrum and CMB TT-spectrum of the emergent scenario and the slow expansion scenario. These figures show that both of these spectra for the slow expansion scenario are the same as that for $\Lambda$CDM, and the spectra of the emergent scenario are lower than that for $\Lambda$CDM at large scales.
\end{abstract}

\maketitle

\section{Introduction}

Inflation~\cite{Guth1981, Linde1982, Albrecht1982} posits an epoch very early in the universe, during which the scale factor grows exponentially with time. It can solve most of problems in the standard cosmology. The primordial scalar perturbations originating from quantum fluctuations during the inflationary epoch not only explain the cosmic microwave background radiation anisotropy but also seed the large-scale structure of the universe~\cite{Mukhanov1981, Lewis2000, Bernardeau2002}. Although it achieves great success, it still suffers from the big bang singularity problem. To solve this intractable problem, some scenarios had been proposed and suggested to construct non-singular or past eternal cosmological models, such as the emergent scenario~\cite{Ellis2004a, Ellis2004b}, the slow expansion scenario~\cite{Piao2003}, the pre-big bang~\cite{Lidsey2000, Gasperini2003}, the cyclic scenario~\cite{Khoury2001, Steinhardt2002, Khoury2004}, the bouncing universe~\cite{Molina-Paris1999, Peter2002} and the other nonsingular model~\cite{Starobinsky1980, Mukhanov1981}.

In the emergent scenario, the universe is assumed to start from an Einstein static universe and then evolves into an inflationary era~\cite{Ellis2004a, Ellis2004b}. Since the universe stems from an Einstein static universe in the emergent scenario, the big bang singularity is avoided. In addition, the e-folding number and the nearly scale-invariant spectral index can also be produced by the inflation of the emergent scenario~\cite{Ellis2004b}. Thus, the emergent scenario has drew lots of attention after it was proposed~\cite{Campo2007, Wu2010, Cai2012, Zhang2014, HuangQ2015, Shabani2017, Shabani2019, HuangQ2020, Khodadi2016, Heydarzade2016, Khodadi2018, Labrana2019, Li2019, Bengochea2021, Ilyas2021, Khodadi2022}. For another model, namely the slow expansion scenario, the universe originates from an Einstein static universe and then enters into an epoch in which the universe expands very slowly. During this epoch, the nearly scale-invariant primordial power spectrum is provided~\cite{Piao2003}. After this scenario was proposed, it has been investigated in lots of modified gravitational theories~\cite{Piao1, Piao2, Piao3, Liu, Liu1, Cai2016, HuangQ2019}. In the slow expansion scenario, the universe reheats after the slow expansion scenario ends. The reheating mechanism had been studied in Ref.~\cite{Liu}. Then, the evolution of the hot big bang cosmology begins. In addition, it was found that in the theory with nonminimal derivative coupling~\cite{Cai2016} and the scalar-tensor theory~\cite{HuangQ2019}, general relativity is recovered and the universe can evolve with the standard cosmology when the slow expansion ends. Similar to the emergent scenario, the big bang singularity is also avoided since the universe originates from an Einstein static universe. Since both the emergent scenario and the slow expansion scenario assume that the universe originates from an Einstein static universe and that a stable Einstein static universe must be stable against both the scalar perturbations and the tensor perturbations, to find a stable Einstein static universe becomes a crucial issue. Fortunately, it was found that a stable Einstein static universe exists in Mimetic gravity~\cite{HuangQ2020}, scalar-fluid theory~\cite{Bohmer2015}, non-minimal derivative coupling model~\cite{Huang2018a, Huang2018b}, braneworld model~\cite{Zhang2016}, Jordan-Brans-Dicke theory~\cite{Huang2014}, Eddington-inspired Born-Infeld theory~\cite{Li2017}, hybrid metric-Palatini gravity~\cite{Bohmer2013}, GUP theory~\cite{Atazadeh2017}, f(R,T) gravity~\cite{Sharif2019}, f(R,T,Q) gravity~\cite{Sharif2018}, massive gravity~\cite{Li2019}, and so on. Thus, the big bang singularity can be solved in the theories of modified gravity by using the emergent scenario and the slow expansion scenario.

It is notable that, except for avoiding the big bang singularity, both the emergent scenario and the slow expansion scenario can produce a nearly scale-invariant primordial power spectrum, which can explain the cosmic microwave background (CMB) radiation anisotropy observed today and provide seeds for the large-scale structure of the observable Universe~\cite{Lewis2000, Bernardeau2002}. CMB observations indicate that there exists a suppression of CMB TT-spectrum at large scales, which was first detected by COBE~\cite{Smoot1992} and recently confirmed by Planck 2018~\cite{Planck2020}. This might correspond to the physics at the very earliest universe. To explain the suppression of CMB TT-spectrum at large scales, several approaches are proposed. One approach is to introduce the spatial curvature in the inflationary model~\cite{Bonga2016, Handley2019}. Another approach is to construct some new models, such as, double inflation~\cite{Feng2003}, hybrid new inflation~\cite{Kawasaki2003}, pre-inflation~\cite{Cicoli2014, Cai2015}, emergent universe~\cite{Labrana2015}, pre-inflationary bounce~\cite{Cai2018}, non-flat XCDM inflation model~\cite{Ooba2018}, warm inflation~\cite{Arya2018}, and so on. Recently, by assuming the Einstein static state as a superinflating phase~\cite{Labrana2015} or a static state phase~\cite{HuangQ2022a, HuangQ2022b}, the CMB TT-spectrum of the emergent scenario was studied in the framework of general relativity, and the results show that the CMB TT-spectrum is suppressed at large scales. However, it is quite unclear that whether the CMB TT-spectrum of the slow expansion scenario is also suppressed at large scales, and whether the CMB TT-spectrum can be utilized to discriminate the emergent scenario from the slow expansion scenario. To answer these questions, we will study the CMB TT-spectrum of the emergent scenario and the slow expansion scenario in the scalar-tensor theory of gravity.

The paper is organized as follows. In section II, we briefly review the field equations of the scalar-tensor theory of gravity. In section III, we study the stability conditions of the Einstein static universe. In section IV, we will give the derivation of equations of motion for perturbations. In section V, we plot the primordial power spectrum and the CMB TT-spectrum of the emergent scenario and the slow expansion scenario. Finally, our main conclusions are shown in Section VI.

\section{Field equations}

In this paper, we consider the following scalar-tensor theory of gravity, whose action takes the following form~\cite{Bergmann,Nordtvedt,Wagoner1970}
\bea\label{action}
S=\int d^4 x \sqrt{-g}\Big[\frac{1}{2}f(\varphi)R-\frac{1}{2}\omega(\varphi)g^{\alpha\beta} \nabla_{\alpha}\varphi \nabla_{\beta}\varphi-V(\varphi)+L_{m}\Big],
\eea
where $R$ is the Ricci curvature scalar, $\varphi$ is the scalar field of scalar-tensor theory, $f(\varphi)$ and $\omega(\varphi)$ are coupling functions of scalar field, $V(\varphi)$ is the potential, and $L_{m}$ represents the Lagrangian density of a perfect fluid. Here, the coupling function $f(\varphi)$ needs to be positive for the gravitons to carry positive energy.

Varying the action~(\ref{action}) with respect to the metric tensor $g^{\alpha\beta}$ and the scalar field $\varphi$, we obtain
\bea\label{f1}
f(R_{\alpha\beta}-\frac{1}{2}g_{\alpha\beta} R)-\nabla_{\alpha}\nabla_{\beta}f+g_{\alpha\beta}\nabla_{\sigma}\nabla^{\sigma}f-\omega\Big(\nabla_{\alpha}\varphi \nabla_{\beta}\varphi-\frac{1}{2}g_{\alpha\beta}\nabla^{\sigma}\varphi \nabla_{\sigma}\varphi\Big)+g_{\alpha\beta}V=T_{\alpha\beta},
\eea
and
\bea\label{f2}
f_{\varphi} R+\omega_{\varphi}\nabla^{\sigma}\varphi \nabla_{\sigma}\varphi+2\omega \nabla_{\sigma}\nabla^{\sigma}\varphi-2V_{\varphi}=0,
\eea
where $f_\varphi$ denotes $\frac{df}{d\varphi}$.

The field equations~(\ref{f1}) can be expressed as the standard form of general relativity $G_{\alpha\beta}=T_{\alpha\beta}^{eff}$ with a modification in the energy-momentum tensor
\bea\label{T}
T_{\alpha\beta}^{eff}=\frac{1}{f}\Big[\nabla_{\alpha}\nabla_{\beta}f-g_{\alpha\beta}(\nabla_{\sigma}\nabla^{\sigma}f+V)+\omega\Big(\nabla_{\alpha}\varphi \nabla_{\beta}\varphi-\frac{1}{2}g_{\alpha\beta}\nabla^{\sigma}\varphi \nabla_{\sigma}\varphi\Big)+T_{\alpha\beta}\Big].
\eea

We consider a homogeneous and isotropic universe described by FLRW metric
\bea\label{m1}
ds^2=-a^2(\eta)dt^2+a^2(\eta)\gamma_{ij}dx^i dx^j, \quad \gamma_{ij}dx^{i} dx^{j}=\frac{dr^{2}}{1-K r^{2}}+r^{2}(d\theta^{2}+\sin^{2}\theta d\varphi^{2}),
\eea
where $K=1,0,-1$ corresponds to a closed, flat and open universe, respectively. The background equation can be obtained by substituting this metric into the field equations~(\ref{f1}), the $0-0$ component gives the Friedmann equation
\bea\label{00}
3\mathcal{H}^2+3K+3\mathcal{H}\frac{f'}{f}=\frac{1}{f}\big(\frac{1}{2}\omega\varphi'^{2}+a^{2}V+a^{2}\rho\big),
\eea
and the $i-i$ component gives
\bea\label{ii}
2\mathcal{H}'+\mathcal{H}^2+\frac{f''}{f}+\mathcal{H}\frac{f'}{f}+K=-\frac{1}{f}\big(\frac{1}{2}\omega \varphi'^{2}-a^{2}V+a^{2}p\big),
\eea
where $'$ denotes a derivative with respect to the conformal time $\eta$, and $\rho$ and $p$ denote the energy density and pressure of the perfect fluid with $p=(\gamma-1)\rho$. Combining Eqs.~(\ref{00}) and ~(\ref{ii}), and eliminating $\rho$, one obtain
\bea\label{tH}
2\mathcal{H}'+\frac{f''}{f}+(3\gamma-2)\Big(\mathcal{H}^2+K+\mathcal{H}\frac{f'}{f}\Big)=\frac{1}{f}\Big[\frac{1}{2}(\gamma-2)\omega \varphi'^{2}+\gamma a^{2}V\Big].
\eea

The field equation~(\ref{f2}) can be expressed as
\bea\label{scalarH}
\varphi''+2\mathcal{H} \varphi'+\frac{1}{2\omega}\big[\omega_{\varphi} \varphi'^{2}+2a^{2}V_{\varphi}-6(\mathcal{H}'+\mathcal{H}^{2}+K)f_{\varphi}\big]=0.
\eea

\section{Einstein static universe}

Both in the emergent scenario and the slow expansion scenario, the universe stems from an Einstein static universe. However, it was found that the Einstein static solution is unstable in scalar-tensor theory of gravity when the perfect fluid is pressureless matter($\gamma=1$) or radiation($\gamma=\frac{4}{3}$) ~\cite{Miao2016}. So, to find a stable Einstein static solution becomes crucial for the emergent scenario and the slow expansion scenario in this theory. In order to find a stable Einstein static solution, we reanalyze the stability of the Einstein static solution by considering $0 \leq \gamma \leq 2$ in scalar-tensor theory of gravity.

\subsection{Static solutions}

The Einstein static solution requires $a=a_{0}=constant$ and $a'_{0}=a''_{0}=0$ which indicates $\mathcal{H}_{0}=\mathcal{H}'_{0}=0$. And Eqs.~(\ref{00}) and ~(\ref{ii}) indicate that a constant $a$ requires $\varphi=\varphi_{0}=constant$ and $\varphi'_{0}=\varphi''_{0}=0$. Thus, for the Einstein static solution, Eqs.~(\ref{tH}) and ~(\ref{scalarH}) show
\bea\label{ESa}
\frac{K}{a_{0}^{2}}=\frac{\gamma V_{0}}{(3\gamma-2)f_{0}},
\eea
and
\bea
\frac{K}{a_{0}^{2}}=\frac{V_{0\varphi}}{3f_{0\varphi}},
\eea
where $_0$ denotes the corresponding static state value, and $_{0\varphi}=\Big(\frac{d}{d\varphi}\Big)_{\varphi=\varphi_{0}}$. In addition, the energy density $\rho_{0}$ is given by Eq.~(\ref{00})
\bea
\rho_{0}=\frac{2V_{0}}{3\gamma-2}.
\eea

The existence conditions of Einstein static solutions require $a^{2}_{0}>0$ and $\rho^{2}_{0}>0$ which mean
\bea\label{ec1}
f_{0}>0, \quad V_{0}>0, \quad \frac{2}{3}<\gamma\leq 2, \quad \frac{V_{0\varphi}}{f_{0\varphi}}>0,
\eea
for $K=1$ and
\bea
f_{0}<0, \quad V_{0}>0, \quad \frac{2}{3}<\gamma\leq 2, \quad \frac{V_{0\varphi}}{f_{0\varphi}}<0.
\eea
for $K=-1$.

Since a stable Einstein static universe is required to be stable against both scalar perturbations and tensor perturbations, we will analyze the stability in the following subsections.

\subsection{Tensor perturbations}

Since tensor perturbations are easy to analyze, we will analyze it at first and the perturbed metric is given as~\cite{Bardeen1980}
\beq\label{PT}
ds^{2}=-a^{2}(\eta)d\eta^{2}+a^{2}(\eta) (\gamma_{ij}+2h_{ij})dx^{i} dx^{j}.
\eeq
Performing a harmonic decomposition for the perturbed variable $h_{ij}$, we obtain
\beq
h_{ij}=H_{T,klm}(t)Y_{ij,klm}(\theta^{n}).
\eeq
Because the quantum numbers $m$ and $l$ do not enter the perturbed differential equations, the harmonic function $Y_{k}=Y_{klm}(\theta^{n})$ satisfies~\cite{Harrison1967}
\bea\label{Yn}
\Delta Y_{k}=-\mathcal{K}^{2}Y_{k}=\Bigg\{
\begin{array}{rrr}
-k(k+2)Y_{k}, \quad k=0,1,2,..., \quad K=+1\\
-k^{2}Y_{k}, \quad\quad\quad\quad k^{2}\geq 0, \quad\quad\quad\quad K=0\\
-(k^{2}+1)Y_{k}, \quad\quad k^{2}\geq 0, \quad\quad\quad K=-1
\end{array}
\eea
where $\Delta$ represents the three-dimensional spatial Laplacian operator. Following Ref.~\cite{Barrow1993, Barrow1995}, considering the static conditions and then substituting the perturbed metric~(\ref{PT}) into the field equations~(\ref{f1}), the equation of tensor perturbations becomes
\beq
H''_{T}+(k^{2}+2K)H_{T}=0.
\eeq
According to this equation, we can find that $k^{2}+2K>0$ must be satisfied for any $k$ to obtain a stable solution. As a result, the Einstein static solutions are stable against the tensor perturbations for the case $K=1$.

\subsection{Scalar perturbations}

Since the Einstein static solutions can be stable against the tensor perturbation for $K=1$, we will analyze the stability of the static solutions under the scalar perturbations in the closed spacetime in this subsection. To achieve this goal, we take the perturbed metric in the Newtonian gauge~\cite{Bardeen1980}
\beq\label{PS}
ds^{2}=-a(\eta)^{2}(1-2\Psi)d\eta^{2}+a(\eta)^{2}(1+2\Phi)\gamma_{ij}dx^{i} dx^{j},
\eeq
where $\Psi$ denotes the Bardeen potential, and $\Phi$ represents the perturbation to the spatial curvature. Similar to the tensor perturbations, considering the static conditions and substituting the perturbed metric~(\ref{PS}) into the field equations~(\ref{f1}) and ~(\ref{f2}), the equation of scalar perturbations are obtained~\cite{Miao2016}
\bea
&& \Psi_{k}-\Phi_{k}=\frac{f_{0\varphi}}{f_{0}}\delta\varphi_{k},\\
&& -2\mathcal{K}^{2}(\Phi_{k}-\Psi_{k})-6\Phi''=\frac{2f_{0\varphi}}{f_{0}}\mathcal{K}^{2}\delta\varphi_{k}
+\frac{3a^{2}_{0}(\gamma-1)\rho_{0}}{f_{0}}\Big[\frac{(2\mathcal{K}^{2}-6)f_{0}}{a^{2}_{0}\rho_{0}}\Phi_{k}+\frac{\mathcal{K}^{2}f_{0\varphi}}{a^{2}_{0}\rho_{0}}\delta\varphi_{k}\Big]\nonumber\\
&& \qquad +\frac{6a^{2}_{0}[(\gamma-1)\rho_{0}-V_{0}]}{f_{0}}\Phi_{k}+\frac{3f_{0\varphi}}{f_{0}}\delta\varphi_{k}''-\frac{3a^{2}_{0}[(\gamma-1)\rho_{0}f_{0\varphi}-V_{0}f_{0\varphi}+f_{0}V_{0\varphi}]}{f_{0}^{2}}\delta\varphi_{k},\\
&& 3\Phi_{k}''+(\mathcal{K}^{2}-6)\Phi_{k}-\frac{f_{0\varphi}}{f_{0}}\mathcal{K}^{2}\delta\varphi_{k}-\frac{\omega_{0}}{f_{0\varphi}}\delta\varphi_{k}''
-\frac{\omega_{0}}{f_{0\varphi}}\mathcal{K}^{2}\delta\varphi_{k}+\frac{a^{2}_{0}f_{0\varphi\varphi}}{f_{0\varphi}^{2}}V_{0\varphi}\delta\varphi_{k}\nonumber\\
&&\qquad \qquad \qquad \qquad \qquad \qquad \qquad \qquad \qquad \qquad \qquad \quad -\frac{a^{2}_{0}}{f_{0\varphi}}V_{0\varphi\varphi}\delta\varphi_{k}=0,\\
&& \Big(\frac{2\omega_{0}f_{0}}{f_{0\varphi}}+3f_{0\varphi}\Big)(\delta\varphi_{k}''+\mathcal{K}^{2}\delta\varphi_{k})
-2a^{2}_{0}\Big(V_{0\varphi}-\frac{f_{0}}{f_{0\varphi}}V_{0\varphi\varphi}+\frac{f_{0}f_{0\varphi\varphi}}{f_{0\varphi}^{2}}V_{0\varphi}\Big)\delta\varphi_{k}\nonumber\\
&&\qquad \qquad \qquad \qquad \quad +\mathcal{K}^{2}(3\gamma-4)f_{0\varphi}\delta\varphi_{k}+2f_{0}(3\gamma-4)(\mathcal{K}^{2}-3)\Phi_{k}=0.
\eea
Then, combining above equations, we obtain two independent equations, which can be written as
\bea
\Bigg(
\begin{array}{r}
\xi_{k}''\\
\Phi_{k}''
\end{array}
\Bigg)
+\mathbf{N}
\Bigg(
\begin{array}{r}
\xi_{k}\\
\Phi_{k}
\end{array}
\Bigg)
=0,
\eea
where $\xi_{k}$ is defined as $\xi_{k}=\delta\varphi_{k}/f_{0}$, and $\mathbf{N}$ is a constant coefficient matrix, which is given as
\bea
\mathbf{N}=\Bigg(
\begin{array}{rr}
N_{11} \quad N_{12}\\
N_{21} \quad N_{22}
\end{array}
\Bigg)
\eea
where
\bea
&&N_{11}=\frac{[2\omega_{0}f_{0}+(3\gamma-1)f^{2}_{0\varphi}]\mathcal{K}^{2}+2a^{2}_{0}f_{0}V_{0\varphi\varphi}-6f_{0}f_{0\varphi\varphi}-6f^{2}_{0\varphi}}{2\omega_{0}f_{0}+3f^{2}_{0\varphi}},\\
&&N_{12}=\frac{2(3\gamma-4)(\mathcal{K}^{2}-3)f_{0\varphi}}{2\omega_{0}f_{0}+3f^{2}_{0\varphi}},\\
&&N_{21}=\frac{[(\gamma-2)\omega_{0}f_{0}f_{0\varphi}-f^{3}_{0\varphi}]\mathcal{K}^{2}+(3f_{0\varphi\varphi}-a^{2}_{0}V_{0\varphi\varphi}-2\omega_{0})f_{0}f_{0\varphi}}{2\omega_{0}f_{0}+3f^{2}_{0\varphi}},\\
&&N_{22}=\frac{[f^{2}_{0\varphi}+2(\gamma-1)\omega_{0}f_{0}]\mathcal{K}^{2}+2(2-3\gamma)\omega_{0}f_{0}-6f^{2}_{0\varphi}}{2\omega_{0}f_{0}+3f^{2}_{0\varphi}}.
\eea
The stability of the static solutions against the scalar perturbations are determined by the eigenvalues of the matrix $\mathbf{N}$, which are expressed as
\bea
\mu^{2}_{1,2}=\frac{M\pm \sqrt{N}}{2},
\eea
with
\bea
M=N_{11}+N_{22}, \quad N=(N_{11}+N_{22})^{2}+4N_{12}N_{21}-4N_{11}N_{22}.
\eea
If the imaginary components of $\mu_{1}$ and $\mu_{2}$ are nonzero, the corresponding Einstein static solutions are unstable. So, the stability conditions can be rewritten as
\bea\label{sc1}
M>0, \quad N>0, \quad M^{2}-N>0.
\eea

Since the homogeneous scalar perturbation corresponds to the case $\mathcal{K}^{2}=k(k+2)=0$ and the inhomogeneous ones correspond to the other case, the stable Einstein static solutions are required to be stable for all values of $k$. So, by considering the existence conditions of Einstein static solution Eq.~(\ref{ec1}) and solving the inequalities ~(\ref{sc1}), we find the Einstein static solutions can be stable under the conditions $f_{0}>0, \omega_{0}<0, V_{0}>0$ with
\bea
&&V_{0\varphi}<0, \frac{19+\sqrt{201}}{24}<\gamma\leq\frac{7}{5}, -\sqrt{-\frac{(2-3\gamma)^{2}f_{0}\omega_{0}}{18(\gamma-1)}}<f_{0\varphi}<-\frac{(2-9\gamma)f_{0}\omega_{0}}{12\gamma-1}, A<V_{0\varphi\varphi}<B,\label{SSC1}\nonumber\\
\\
&&V_{0\varphi}>0, \frac{19+\sqrt{201}}{24}<\gamma\leq\frac{7}{5}, \frac{(2-9\gamma)f_{0}\omega_{0}}{12\gamma-1}<f_{0\varphi}<\sqrt{-\frac{(2-3\gamma)^{2}f_{0}\omega_{0}}{18(\gamma-1)}},A<V_{0\varphi\varphi}<B,\label{SSC2}\nonumber\\
\eea
where
\bea
&&A=\frac{[3f_{0}f_{0\varphi\varphi}+(12-9\gamma)f^{2}_{0\varphi}+(2-3\gamma)f_{0}\omega_{0}]V_{0\varphi}}{3f_{0}f_{0\varphi}}+\sqrt{\frac{(4-3\gamma)^{2}(2f_{0}\omega_{0}+3f^{2}_{0\varphi})V^{2}_{0\varphi}}{3f^{2}_{0}}},\nonumber\\
&&B=\frac{V_{0\varphi}f_{0\varphi\varphi}}{f_{0\varphi}}+\frac{2f_{0\varphi}V_{0\varphi}}{(3\gamma-2)f_{0}}.\nonumber
\eea
As we will see in the next section, to avoid the ghost and gradient instabilities, $f_{0\varphi}>\sqrt{-\frac{2}{3}f_{0}\omega_{0}}$ (Eq.~(\ref{ngg})) should be satisfied. As a result, the stability conditions ~(\ref{SSC1}) is excluded, and the Einstein static solutions can be stable under the conditions ~(\ref{SSC2}).

\begin{figure*}[htp]
\begin{center}
\includegraphics[width=0.45\textwidth]{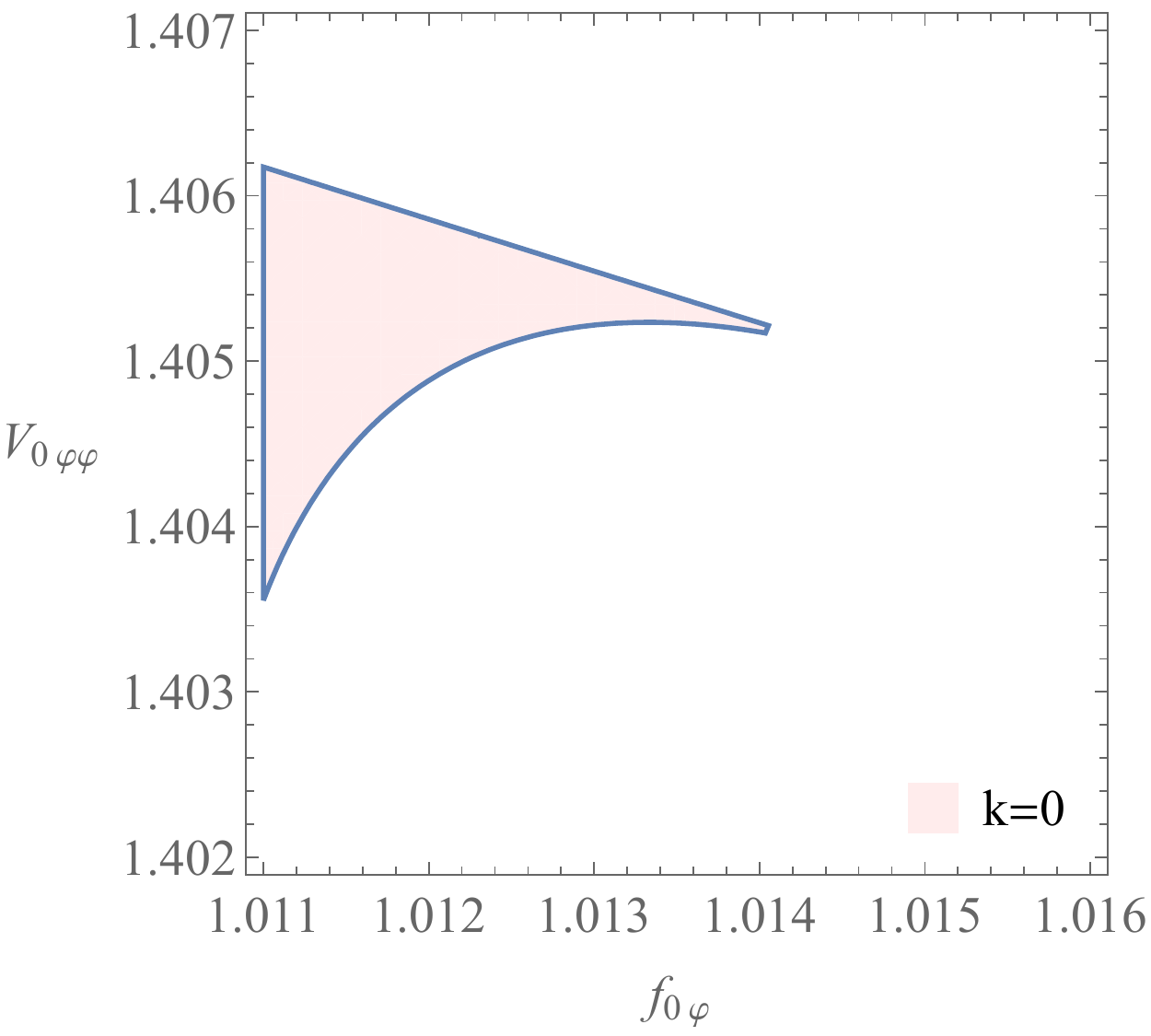}
\includegraphics[width=0.45\textwidth]{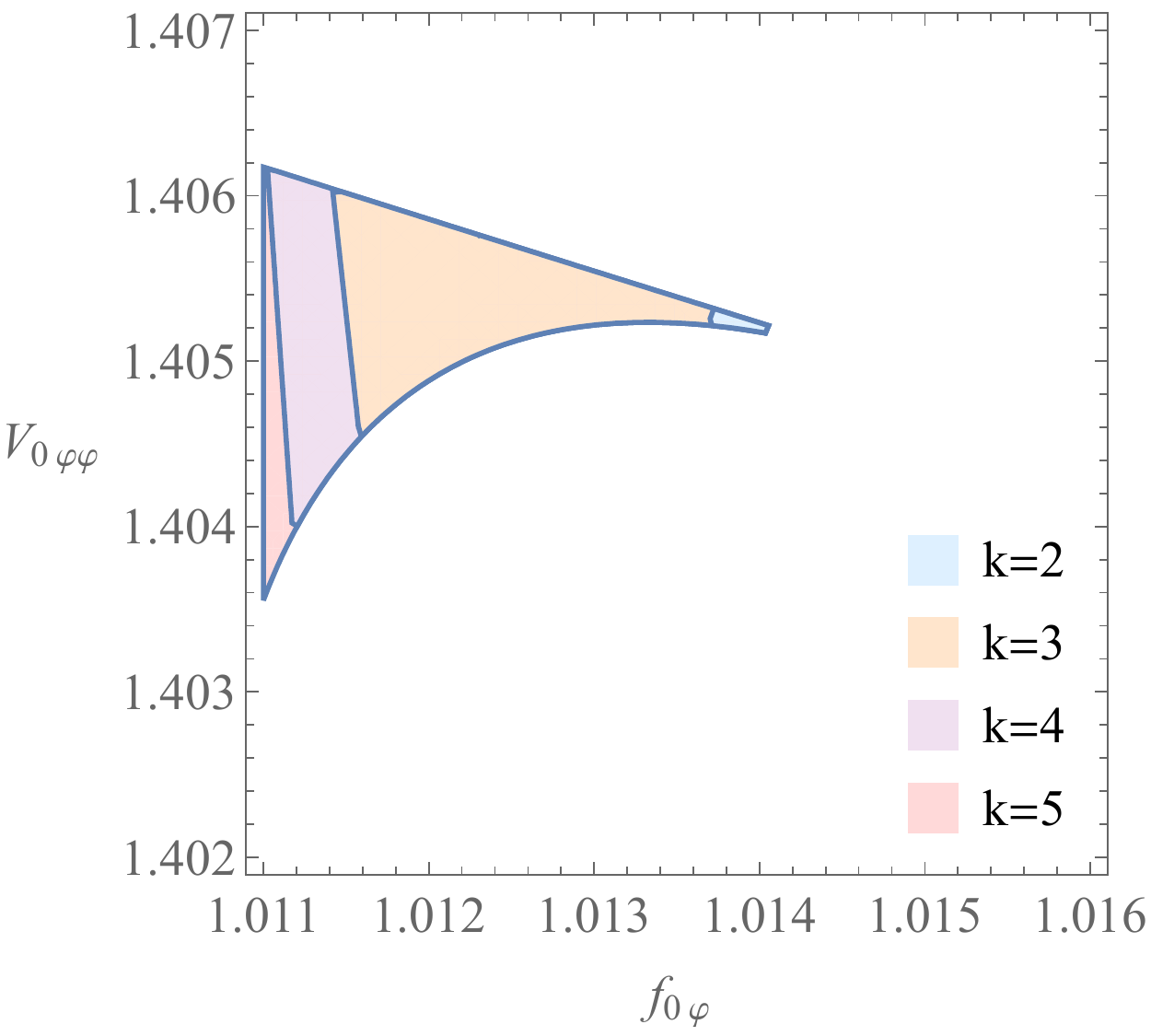}
\caption{\label{Fig1} Stability regions in $(V_{0\varphi\varphi},f_{0\varphi})$ plane under homogeneous and inhomogeneous scalar perturbations. $k=0$ represents the homogeneous scalar perturbations while $k=2,3,4,5$ correspond to the inhomogeneous ones. These figures are plotted for $f_{0}=1.532$, $\omega_{0}=-1$, $V_{0\varphi}=0.9052$, $\gamma=1.4$ and $f_{0\varphi\varphi}=0.964$.}
\end{center}
\end{figure*}

\begin{figure*}[htp]
\begin{center}
\includegraphics[width=0.45\textwidth]{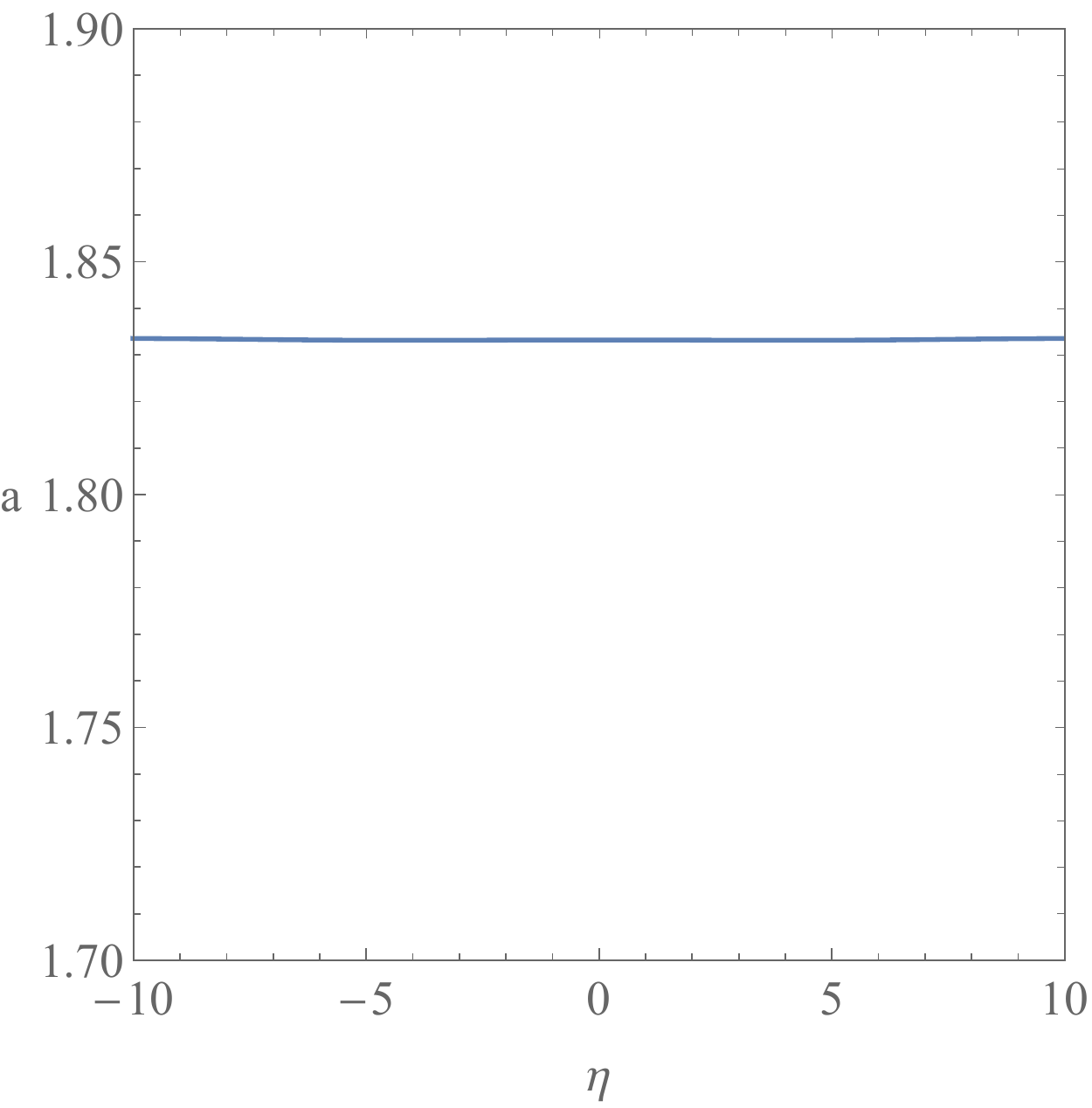}
\includegraphics[width=0.45\textwidth]{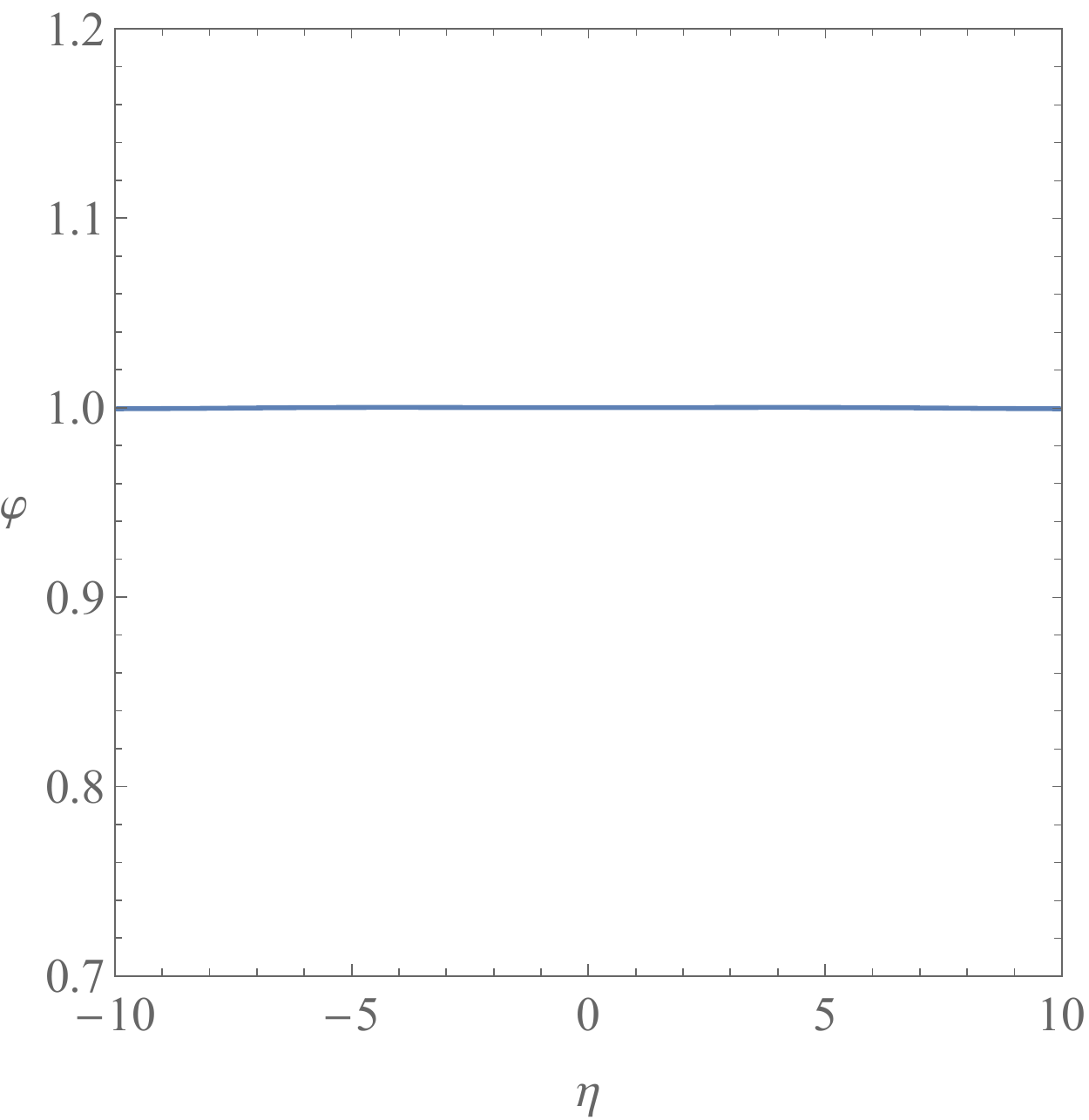}
\caption{\label{Fig2} Evolutionary curves of scale factor $a$ and scalar field $\varphi$ under the stability conditions. These figures are plotted for $f=1+0.05\varphi+0.482\varphi^{2}$, $\omega=-\frac{1}{\varphi}$, $V=0.5138-0.5\varphi+0.7026\varphi^{2}$ and $\gamma=1.4$.}
\end{center}
\end{figure*}

From the stability conditions ~(\ref{SSC2}), we can see that the range of value $\gamma$ is extremely narrow. To make the stability conditions more intuitive, we have plotted contours of the stable regions of the homogeneous and inhomogeneous scalar perturbations in Fig.~(\ref{Fig1}) by considering $f=1+\xi \varphi+\lambda \varphi^{2}$, $\omega=-\frac{w}{\varphi}$ and $V=V_{1}+\alpha \varphi+\beta \varphi^{2}$. To make sure the value of $f_{0}$, $\omega_{0}$, $V_{0\varphi}$, $\gamma$ and $f_{0\varphi\varphi}$ satisfy the stability conditions ~(\ref{SSC2}), we take $\xi=0.05$, $\lambda=0.482$, $w=1$, $V_{1}=0.5138$, $\alpha=-0.5$, $\beta=0.7062$ and $\gamma=1.4$ as an example. In this figure, $k=0$ denotes the homogeneous scalar perturbations, and $k=2,3,4,5$ correspond to the inhomogeneous ones. For the inhomogeneous scalar perturbations, the stable region becomes larger and larger with the increase of $k$, and the stable region of $k=0$ overlaps with the case $k=5$. As a result, the region for $k=2$ is the smallest stable region which represents the stable region for the Einstein static solutions. Thus, a stable Einstein static universe can exist in the scalar-tensor theory of gravity and an example is depicted in Fig.~(\ref{Fig2}) by solving Eqs.~(\ref{tH}) and ~(\ref{scalarH}) numerically under the stability conditions.

\section{Perturbations}

To analyze the power spectrum of the emergent scenario and the slow expansion scenario in scalar-tensor theory of gravity, we are required to obtain the equation of motion for perturbation. In this section, we will derive this equation.

\subsection{Conformal transformation to Einstein gravity}

The scalar-tensor theory of gravity can be transformed into Einstein gravity by performing a conformal transformation on the metric~\cite{Hwang,Hwang1990,Farese,Qiu,Glavan2015}
\bea
\tilde{g}_{\alpha\beta}=\Omega^2 g_{\alpha\beta},
\eea
and the corresponding action becomes
\bea\label{action1}
\tilde{S}=\int d\tilde{t} d^3 x \sqrt{-\tilde{g}}\big(\frac{1}{2}\tilde{R}-\frac{1}{2}\tilde{\nabla}^\sigma \tilde{\varphi}\tilde{\nabla}_\sigma \tilde{\varphi}-\tilde{V}\big),
\eea
which is the action for a minimal coupling single scalar field $\tilde{\varphi}$ and the conformal factor is $\Omega=\sqrt{f}$, and the corresponding variables are defined as
\bea\label{relation}
d\tilde{t}=\Omega dt, \quad \tilde{a}=\Omega a, \quad \tilde{\Phi}=\Phi+\delta\Omega, \quad \tilde{\Psi}=\Psi-\delta\Omega, \quad d\tilde{\varphi}=\sqrt{\frac{\omega}{f}+\frac{3}{2}\frac{f_{,\varphi}^2}{f^2}} d\varphi, \quad \tilde{V}=\frac{V}{f^2}.
\eea
Here, the Lagrangian density of a perfect fluid $L_{m}$ is ignored for the reasons: (i) The evolution of the universe is dominated by the scalar field before inflation ends. (ii) During the inflationary epoch, the scale factor of the universe increases exponentially, and the energy density of the perfect fluid becomes very small. Thus, the action~(\ref{action}) without the term $L_{m}$ and ~(\ref{action1}) are fully equivalent~\cite{Hwang,Hwang1990,Farese,Qiu,Glavan2015,Weenink,Prokopec2012,Prokopec2013}. We will discuss the cosmological perturbations in action~(\ref{action1}) since it is easier to analyze. The background equation and the equation of motion for the field $\tilde{\varphi}$ can be written as follows
\bea\label{bge1}
3\tilde{H}^2+3\frac{K}{\tilde{a}^2}=\frac{1}{2}\dot{\tilde{\varphi}}^2+\tilde{V},
\eea
\bea\label{bge2}
3\tilde{H}^2+2\dot{\tilde{H}}+\frac{K}{\tilde{a}^2}=-\big(\frac{1}{2}\dot{\tilde{\varphi}}^2-\tilde{V}\big),
\eea
and
\bea
\ddot{\tilde{\varphi}}+3\tilde{H}\dot{\tilde{\varphi}}+\tilde{V}_{\tilde{\varphi}}=0.
\eea
Here, the no ghost condition $\dot{\tilde{\varphi}}^2>0$ gives $\omega f+\frac{3}{2}\big(\frac{df}{d\varphi}\big)^2>0$ which is contained in expression~(\ref{E}), and a dot represents the derivative with respect to the cosmic time $t$.

In order to obtain the perturbation equations with curvature $K$, we use the method developed by Garriga and Mukhanov~\cite{Garriga,Mukhanov2}. Then, the perturbation equations for $0-0$ and $0-i$ components can be written as~\cite{Mukhanov,Garriga}
\bea\label{ee1}
2\Big[\frac{1}{\tilde{a}^2}\Delta\tilde{\Psi}-3\tilde{H}\dot{\tilde{\Psi}}+3\big(\frac{K}{\tilde{a}^2}\tilde{\Psi}-\tilde{H}^2 \tilde{\Phi}\big)\Big]=\dot{\tilde{\varphi}}\dot{\delta\tilde{\varphi}}-\tilde{\Phi}\dot{\tilde{\varphi}}^2+\tilde{V}_{\tilde{\varphi}}\delta\tilde{\varphi},
\eea
\bea\label{ee2}
2\nabla_{i}\big(\dot{\tilde{\Psi}}+\tilde{H}\tilde{\Phi}\big)=\dot{\tilde{\varphi}} \nabla_{i}\delta\tilde{\varphi},
\eea
where $\Delta=\nabla^2$ is Laplace operator, and the component for $i\neq j$ is
\bea
\frac{1}{\tilde{a}^2} \nabla_{j}\nabla^{i} \big(\tilde{\Phi}-\tilde{\Psi}\big)=0,
\eea
which gives $\tilde{\Phi}=\tilde{\Psi}$. After introducing two new variables
\bea
\tilde{\xi}=\frac{2\tilde{a}}{\tilde{H}}\tilde{\Psi}, \\
\tilde{\zeta}=\tilde{\Psi}+\tilde{H}\frac{\delta\tilde{\varphi}}{\dot{\tilde{\varphi}}}-\frac{2K}{\tilde{a}^2 \dot{\tilde{\varphi}}^2}\tilde{\Psi},
\eea
the equations~(\ref{ee2}) and ~(\ref{ee1}) can be simplified as
\bea\label{ee3}
\dot{\tilde{\xi}}=\tilde{a}\frac{\dot{\tilde{\varphi}}^2}{\tilde{H}^2}\tilde{\zeta},
\eea
\bea\label{ee4}
\dot{\tilde{\zeta}}=\frac{\tilde{H}^2}{\tilde{a}^3 \dot{\tilde{\varphi}}^2}\big(\Delta+Y K\big)\tilde{\xi},
\eea
where
\bea\label{ee5}
Y=-2\frac{\tilde{V}_{\tilde{\varphi}}}{\tilde{H}\dot{\tilde{\varphi}}}.
\eea
The detailed process is given in Appendix.

In order to obtain the amplitude of quantum fluctuations, one needs to expand the action for the gravitational and scalar fields to second order in perturbations which are cumbersome. Since the second order perturbation action can be inferred directly from the equations of motion~(\ref{ee3}) and ~(\ref{ee4}), these cumbersome steps can be avoided. The detailed steps are given in Ref.~\cite{Garriga,Mukhanov2}. Thus, the action reproducing the perturbation equations~(\ref{ee3}) and ~(\ref{ee4}) can be written as
\bea\label{action0}
\tilde{S}=\int\Big[\tilde{\xi}\tilde{O}\dot{\tilde{\zeta}}-\frac{1}{2}\frac{\tilde{H}^2}{\tilde{a}^3 \dot{\tilde{\varphi}}^2}\tilde{\xi}\tilde{O}(\Delta+Y K)\tilde{\xi}+\frac{1}{2}\frac{\tilde{a} \dot{\tilde{\varphi}}^2}{\tilde{H}^2}\tilde{\zeta}\tilde{O}\tilde{\zeta}\Big]d\tilde{t} d^3 x,
\eea
where $\tilde{O}=\Delta+3K$ is a time-independent operator. Expressing $\tilde{\xi}$ in terms of $\dot{\tilde{\zeta}}$ by equation~(\ref{ee4}), the action can be reduced to
\bea\label{action2}
\tilde{S}=\int z^2\big[\tilde{\zeta}'^2+\tilde{\zeta}(\Delta+Y K)\tilde{\zeta}\big]d\eta d^3 x,
\eea
where prime denotes the derivative with respect to the conformal time $\eta$, and the variable $z$ is
\bea
z=\frac{\tilde{a} \tilde{\varphi'}}{\mathcal{\tilde{H}}}\sqrt{\frac{\tilde{O}}{\Delta+Y K}}.
\eea
The Laplacian $\Delta$ should be understood as a $c-number$ and represents the corresponding eigenvalue.

\subsection{Equation of motion for perturbation}

The second perturbation action in the scalar-tensor theory of gravity can be obtained by a transformation for the action~(\ref{action2}). Introducing the canonical quantization variable $v=z\tilde{\zeta}$ and utilizing the corresponding transform equations~(\ref{relation}), the action~(\ref{action2}) can be rewritten as follows
\bea\label{action3}
S=\frac{1}{2}\int \Big[v'^2+v(\Delta+Y K)v+\frac{z''}{z}v^2\Big]d\eta d^3 x.
\eea
Here, $z$ and $Y$ are
\bea\label{z}
z=\frac{a^2 \varphi' \sqrt{E}}{(a\sqrt{f})'}\sqrt{\frac{\Delta+3K}{\Delta+Y K}},
\eea
and
\bea\label{Y}
Y=2\frac{(a^2 \varphi' \sqrt{E})'/(a^2 \varphi' \sqrt{E})}{(a\sqrt{f})'/(a\sqrt{f})},
\eea
where
\bea\label{E}
E=f\Big(\omega+\frac{3}{2}\frac{f^2_{\varphi}}{f}\Big).
\eea
To avoid the ghost and gradient instabilities, $f>0$ and $E>0$ should be satisfied. In previous section, we find $f_{0}>0$ and $\omega_{0}<0$ (Eqs.~(\ref{SSC1}) and ~(\ref{SSC2})) are required to obtain a stable Einstein static universe. Under the conditions $f_{0}>0$ and $\omega_{0}<0$, the constraint condition $E>0$ gives
\beq\label{ngg}
f_{0\varphi}>\sqrt{-\frac{2}{3}f_{0}\omega_{0}}.
\eeq
Taking this condition ~(\ref{ngg}) into consideration, the stability conditions ~(\ref{SSC1}) are excluded. Thus, when the conditions ~(\ref{SSC2}) are satisfied, the Einstein static universe is stable and there is no the ghost and gradient instabilities.

Varying the action~(\ref{action3}) with respect to $v$, one can straightforwardly get the equation of motion for the variable as
\bea\label{em}
v''-(\Delta+Y K)v-\frac{z''}{z}v=0.
\eea

Expressing $\tilde{\zeta}$ and $\dot{\tilde{\zeta}}$ in the action~(\ref{action0}) in terms of $\tilde{\dot{\xi}}$ and $\tilde{\xi}$, we obtain
\bea
u''-(\Delta+Y K)u-\frac{Z''}{Z}u=0,
\eea
where
\bea
u=\frac{a \sqrt{f^3}}{\varphi' \sqrt{E}}\tilde{\psi}, \quad Z=\frac{(a\sqrt{f})'}{a^2 \varphi' \sqrt{E}}.
\eea
The equation of motion for the variable $u$ was obtained in Ref.~\cite{Hwang1,Hwang1991} by doing some calculations.

\section{CMB power spectrum}

In order to discuss whether the CMB TT-spectrum can discriminate the emergent scenario from the slow expansion scenario, we study the primordial power spectrum and the CMB TT-spectrum of the emergent scenario and the slow expansion scenario in this section.

\subsection{Emergent scenario}

In emergent scenario, the universe stems from an Einstein static state, and then evolves into an inflationary epoch~\cite{Ellis2004a, Ellis2004b}. To realize this transition, there exist two different approaches: (i) assuming the Einstein static state defined by $a'=0$ and then invoking an instantaneous transition to the inflationary epoch~\cite{Wu2010,HuangQ2015}. (ii) considering the evolution of the scale factor as $a(t)=a_{0}+A e^{H_{0}t}$~\cite{Ellis2004a, Ellis2004b}. In our previous work~\cite{HuangQ2022a}, we found that both approaches produce the same CMB TT-spectra. So, in this section, we will adopt the first approach.

Following Ref.~\cite{HuangQ2022a, Thavanesan2021, Shumaylov2022}, considering the Einstein static conditions, the scale factor in Einstein static state $a_{0}$ is given by Eq.~(\ref{ESa}). During this epoch, the variable $Y$ and $z$ in Eqs.~(\ref{Y}) and ~(\ref{z}) reduce to
\bea
Y \approx 0, \quad z \approx 0.
\eea
So, the equation of motion for the variable $v$ (Eq.~(\ref{em})) can be written as
\bea
v''_{k}+k^{2}_{-} v_{k}=0, \quad k^{2}_{-}=k(k+2),
\eea
which has the solution
\bea\label{vk0}
v_{k}(\eta)=\sqrt{\frac{1}{2k}}e^{-i k_{-} \eta},
\eea
where the normalization conditions $v_{k} v_{k}^{*'}-v^{'}_{k} v_{k}^{*}=i$ and the Bunch-Davies vacuum are considered.

\begin{figure*}[htp]
\begin{center}
\includegraphics[width=0.45\textwidth]{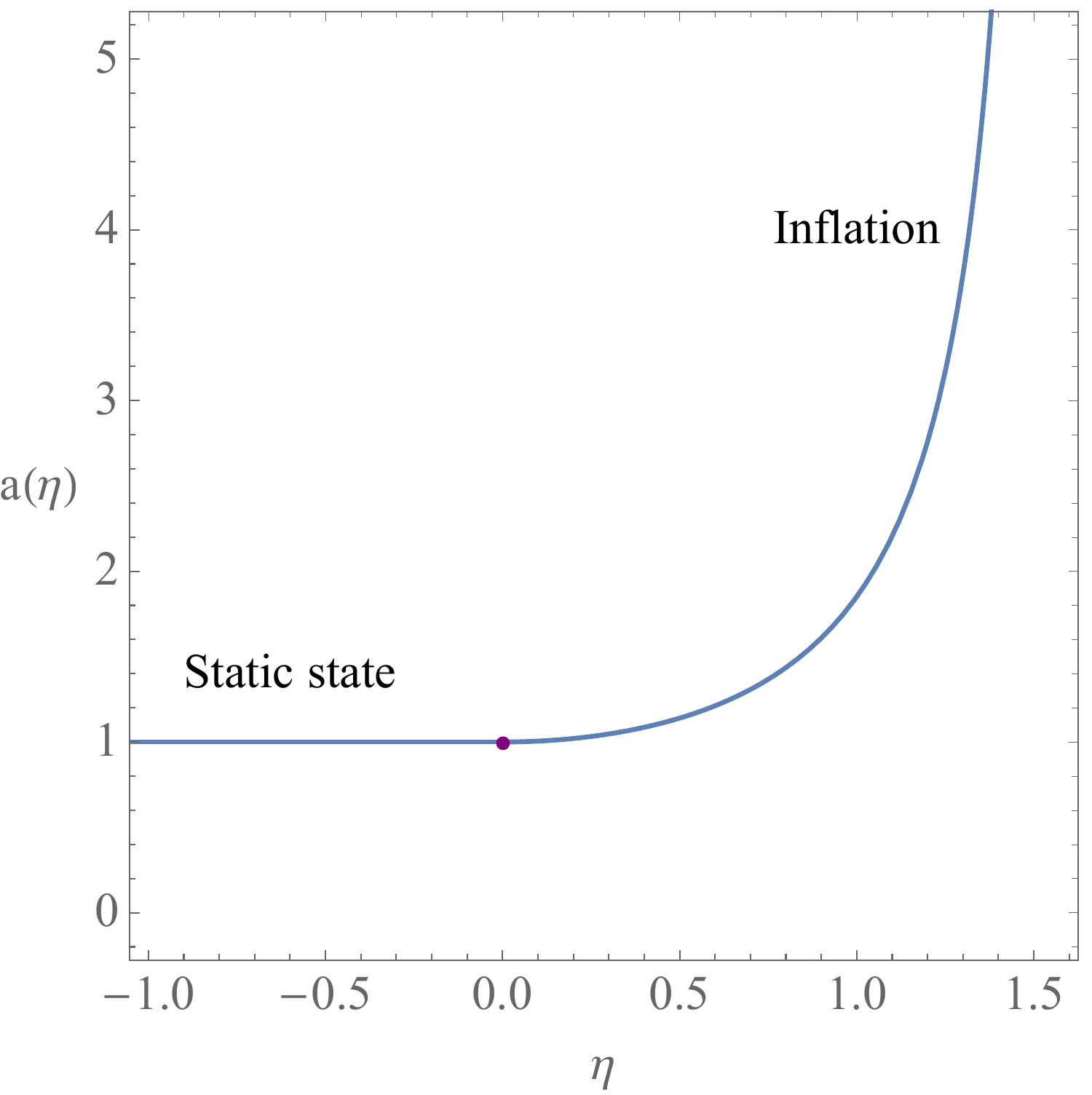}
\caption{\label{Fig3} Evolutionary curve of scale factor $a$. The scale factor in the Einstein static state has been chosen as $a_{0}=1$ and the purple point denotes the transition point.}
\end{center}
\end{figure*}

In slow-roll region, using the slow-roll conditions $f'' \ll \mathcal{H} f' \ll \mathcal{H}^{2} f$ and $\varphi'^{2} \ll a^{2}V$ ~\cite{Barrow1995a}, Eqs.~(\ref{00}) and ~(\ref{ii}) reduces to
\bea
\mathcal{H}'-\mathcal{H}-1 \approx 0,
\eea
which has the solution
\bea
a=\frac{a_{0}}{\cos(\eta-\eta_{t})}.
\eea
Thus, the scale factor for the emergent scenario can be expressed as
\bea
a(\eta)=\Bigg\{
\begin{array}{ll}
a_0,\quad & \eta<\eta_t\\
\frac{a_0}{\cos(\eta-\eta_t)},\quad & \eta_t \leq \eta < \eta_t+\frac{\pi}{2},\label{Comp}
\end{array}
\eea
which is also obtained in general relativity~\cite{HuangQ2022a,HuangQ2022b}. With $\eta$ approaching to $\eta_{t}+\frac{\pi}{2}$, the universe freezes out into the inflationary phase. The evolutionary curve of scale factor $a_{0}$ is shown in Fig.~(\ref{Fig3}), the purple point denotes the transition point. To realize this transition, one can break the stability conditions by considering the scalar potential or the equation of state varying with conformal time $\eta$ slowly~\cite{Wu2010,HuangQ2015,HuangQ2020}. Once $\eta$ evolves to a critical point, the stability conditions will break down automatically.

During the slow-roll region, the variable $Y$ and $z$ reduce to
\bea
Y \approx 4, \quad z \approx \frac{a \varphi' \sqrt{\omega}}{\mathcal{H}}\sqrt{\frac{k(k+2)+3}{k(k+2)+4}}.
\eea
So, the equation of motion for the variable $v$ (Eq.~(\ref{em})) becomes
\bea
v''_{k}+\Big\{k^{2}_{+}-\frac{2}{\big[\eta-(\eta_{t}+\frac{\pi}{2})\big]^{2}} \Big\}v_{k}=0, \quad k^{2}_{+}=k(k+2)-\frac{17}{3},
\eea
which has the solution taking the form
\bea\label{vkcd}
v_k(\eta)=\sqrt{\frac{\pi}{4}}\sqrt{\big(\eta_t+\frac{\pi}{2}\big)-\eta}\Big[C_k H^{(1)}_{3/2}\Big(k_{+}\big((\eta_t+\frac{\pi}{2})-\eta\big)\Big)+ D_k H^{(2)}_{3/2}\Big(k_{+}\big((\eta_t+\frac{\pi}{2})-\eta\big)\Big)\Big],
\eea
where $H^{(1)}$ and $H^{(2)}$ are the Hankel functions of the first and second kinds.

To determine $C_{k}$ and $D_{k}$, we use the continuity condition of $v_{k}$ and $v'_{k}$ to match Eqs.~(\ref{vk0}) and ~(\ref{vkcd}) at the transition time $\eta_{t}$ and obtain
\bea
&& C_k=\frac{1}{4}e^{-i k_{-} \eta_t}\sqrt{\frac{1}{k_{-}}}\Big[i \pi k_{+} H^{(2)}_{1/2}\Big(\frac{\pi}{2}k_{+}\Big)+(-2i+\pi k_{-})H^{(2)}_{3/2}\Big(\frac{\pi}{2}k_{+}\Big)\Big],\\
&& D_k=-\frac{1}{4}e^{-i k_{-} \eta_t}\sqrt{\frac{1}{k_{-}}}\Big[i \pi k_{+} H^{(1)}_{1/2}\Big(\frac{\pi}{2}k_{+}\Big)+(-2i+\pi k_{-})H^{(1)}_{3/2}\Big(\frac{\pi}{2}k_{+}\Big)\Big].
\eea

The curved primordial power spectrum of the comoving curvature perturbation $\mathcal{R}$ is defined as
\bea\label{PR}
\mathcal{P}_{\mathcal{R}}=\frac{k^3}{2\pi^2}\left| \mathcal{R}_k \right|^2 =\frac{k^3}{2\pi^2}\left| \frac{v_k}{z_{k}} \right|^2.
\eea
Then, substituting Eq.~(\ref{vkcd}) into Eq.~(\ref{PR}), we obtain the curved primordial power spectrum of $\mathcal{R}$
\bea
\mathcal{P}_{\mathcal{R}} &&=\frac{k^3}{2\pi^2}\left| \mathcal{R}_k \right|^2 \approx \lim_{\eta\rightarrow\eta_t+\frac{\pi}{2}}
\frac{1}{4\pi^{2}}\frac{1}{\frac{a^{2} \varphi'^{2}\omega}{\mathcal{H}} \frac{k(k+2)+3}{k(k+2)+4}} \frac{k^{3}}{k^{3}_{+}} \frac{1}{\big[\eta-(\eta_{t}+\frac{\pi}{2})\big]^{2}} \left| C_k-D_k \right|^2 \nonumber\\
&&=A_{s} \frac{k^{3}}{k^{3}_{+}} \frac{k(k+2)+3}{k(k+2)+4} \left| C_k-D_k \right|^2,
\eea
where formally diverging parameters are absorbed into the usual scalar power spectrum amplitude $A_{s}$~\cite{Thavanesan2021, Shumaylov2022}. And the analytical primordial power spectrum can be parameterized as
\bea
\mathcal{P}_{\mathcal{R}}=A_{s} \Big(\frac{k}{k_{*}}\Big)^{n_{s}-1} \frac{k^{3}}{k^{3}_{+}} \frac{k(k+2)+3}{k(k+2)+4} \left| C_k-D_k \right|^2,
\eea
where $k_{*}=0.05Mpc^{-1}$ corresponds to the pivot perturbation mode.

\subsection{Slow expansion scenario}

In slow expansion scenario, the universe also stems from an Einstein static state, and then evolves into a slowly expanded epoch which can generate the scale invariant primordial power spectrum~\cite{Piao2003}. In scalar-tensor theory of gravity, by considering $f(\varphi)=1-\xi \varphi+\lambda \varphi^{2}$, $\omega(\varphi)=\omega_{0}\varphi^{-1}$ and $V(\varphi)=-V_{0} \varphi^{\frac{3}{2}}$, the slow expansion scenario was analyzed and it was found that the analytical primordial power spectrum is scale invariant and has the form~\cite{HuangQ2019}
\bea
\mathcal{P}_{\mathcal{R}}=\frac{k^{3}}{2\pi^{2}}\left| \frac{v_{k}}{z_{k}} \right|^2 \approx \frac{V_{0}\xi^{\frac{5}{2}}}{128\pi^{2}}=A_{s}.
\eea
Parameterizing this primordial power spectrum, it can be written as
\bea
\mathcal{P}_{\mathcal{R}}=A_{s} \Big(\frac{k}{k_{*}}\Big)^{n_{s}-1},
\eea
which is the same as that in $\Lambda$CDM model.

\subsection{Power spectrum}

\begin{figure*}[htp]
\begin{center}
\includegraphics[width=0.44\textwidth]{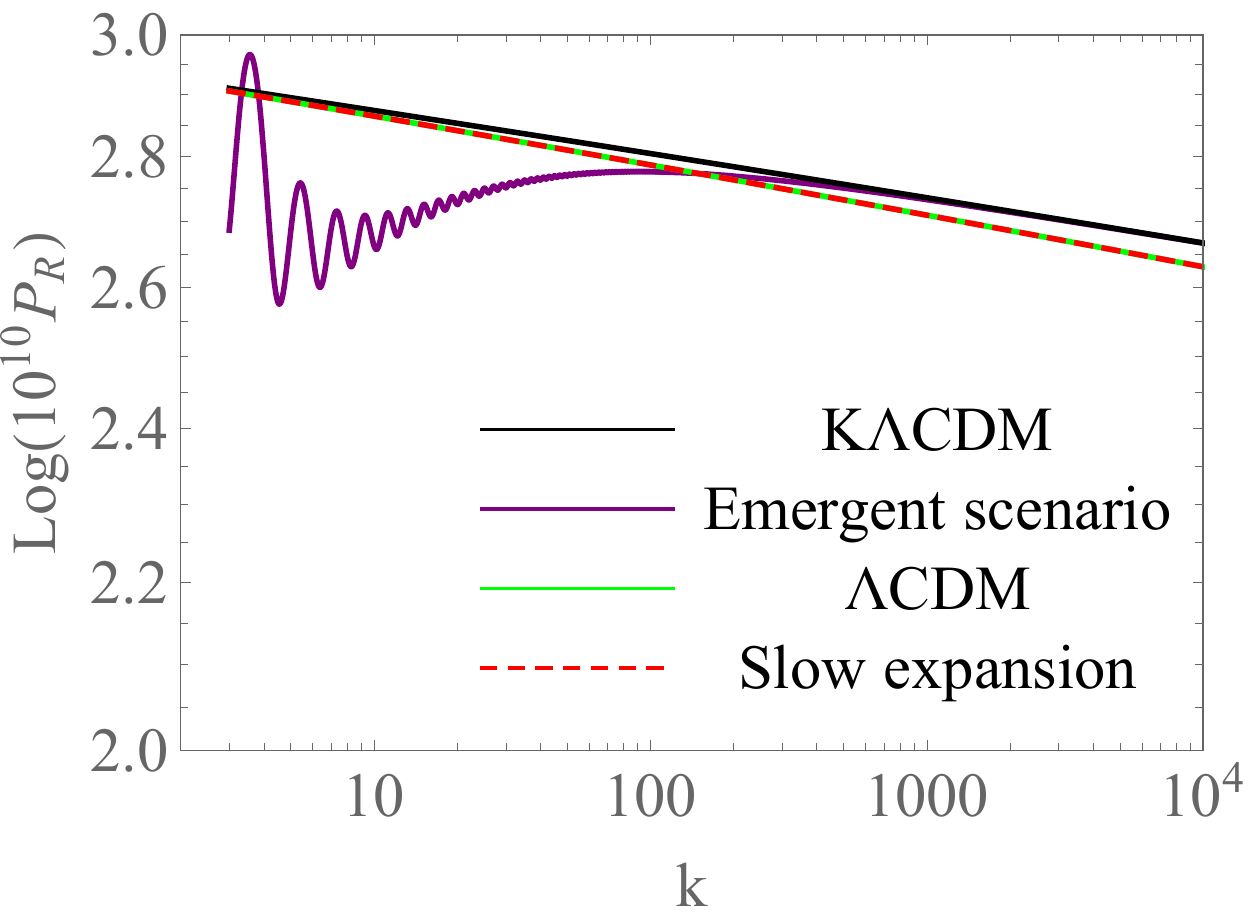}
\includegraphics[width=0.46\textwidth]{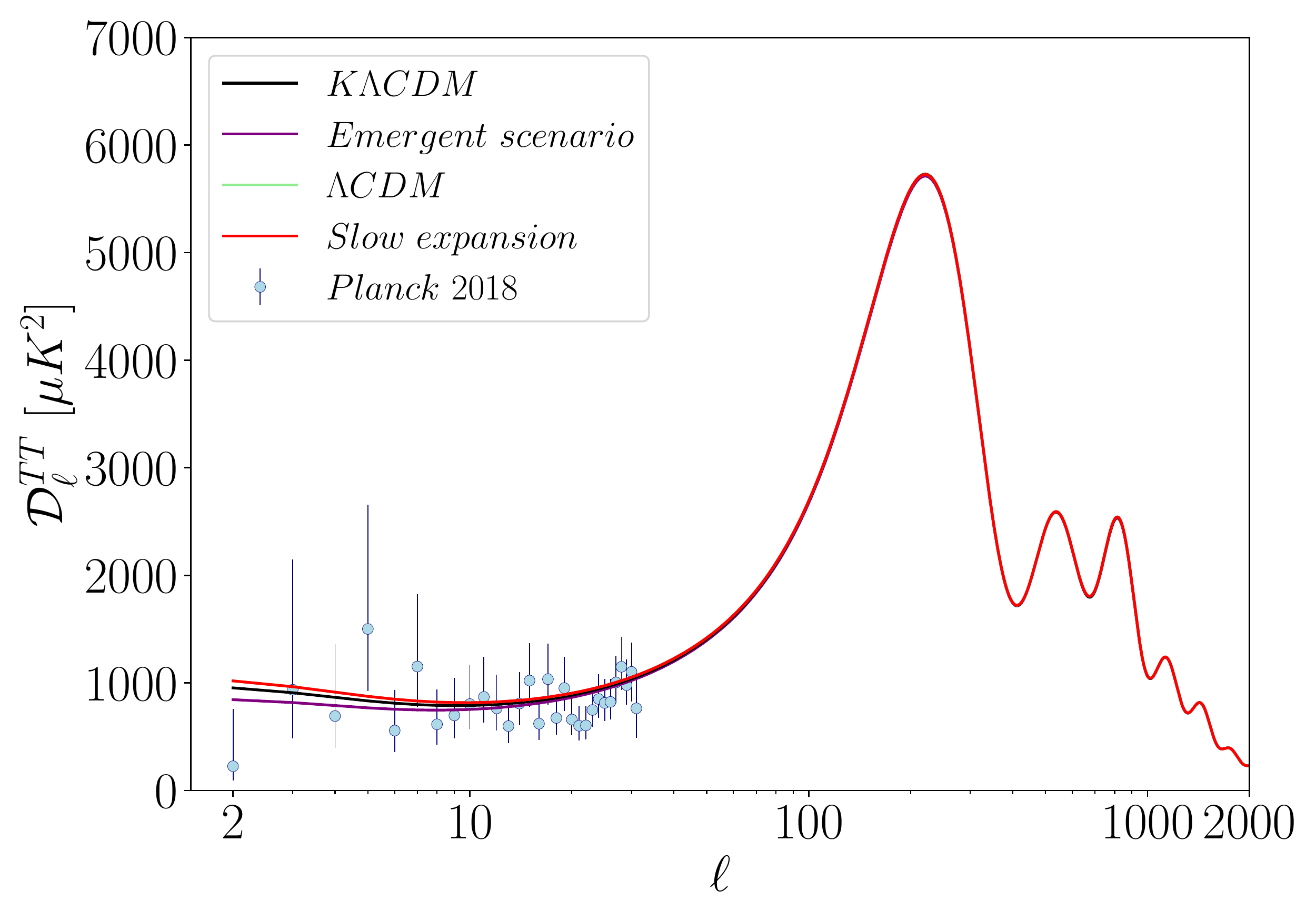}
\caption{\label{Fig4} Primordial power spectrum and CMB TT-spectrum for the emergent scenario and the slow expansion scenario.}
\end{center}
\end{figure*}

To plot the primordial power spectrum in the closed universe, we use the Planck 2018 results in the curved universes best-fit data (TT,TE,EE+lowl+lowE+lensing) $A_s=2.0771 \pm 0.1017 \times 10^{-9}$ and $n_s = 0.9699 \pm 0.0090$. For the flat universe, the Planck 2018 results in Ref.~\cite{Planck2020} is adopted. In the left panel of Fig.~(\ref{Fig4}), we have plotted the primordial power spectrum for $\Lambda$CDM, K$\Lambda$CDM, the emergent scenario and the slow expansion scenario. This figure shows that the primordial power spectra of the slow expansion scenario and $\Lambda$CDM are overlapped. Comparing to $\Lambda$CDM, K$\Lambda$CDM and the slow expansion scenario, the primordial power spectrum of emergent scenario oscillates and is suppressed during the region $k<100$.

Then, using CLASS code~\cite{Blas2011}, we have depicted the CMB TT-spectrum in the right panel of Fig.~(\ref{Fig4}). From this figure, we can see that the CMB TT-spectrum of the slow expansion scenario is the same as that in $\Lambda$CDM, and the CMB TT-spectrum of emergent scenario is lower than the one in $\Lambda$CDM for $l<10$. Thus, comparing to the slow expansion scenario, the emergent scenario can produce a lower CMB TT-spectrum at large scales.

\section{Conclusion}

The scalar-tensor theory is an extension of general relativity by coupling a scalar field $\varphi$ to the Ricci scalar $R$ with terms $f(\varphi)R$, and it can be expressed as general relativity with a modified energy-momentum tensor. In this paper, we study the primordial power spectrum and CMB TT-spectrum of the emergent scenario and the slow expansion scenario in the scalar-tensor theory of gravity. Since both in the emergent scenario and in the slow expansion scenario, the universe stems from an Einstein static universe, we analyze the stability of the Einstein static universe in scalar-tensor theory of gravity at first, and find the Einstein static universe can be stable against both scalar and tensor perturbations under the certain conditions.

Assuming the emergent scenario starts from an Einstein static universe followed by an instantaneous transition to an inflationary phase, we study the primordial power spectrum for the emergent scenario and obtain the analytical approximations of this spectrum. To comparing the primordial power spectrum and CMB TT-spectrum of the emergent scenario and the slow expansion scenario, we have plotted these spectra by using Planck 2018 results. These figures show that both of these spectra for the slow expansion scenario are the same as the one in $\Lambda$CDM, and the spectra of the emergent scenario are lower than the one in $\Lambda$CDM at large scales. Thus, comparing to the slow expansion scenario, the emergent scenario can produce a lower CMB TT-spectrum at large scales.

\section{Appendix}

Equation~(\ref{ee2}) can be written as
\bea
(\tilde{a}\tilde{\Psi})^{.}=\frac{1}{2}\tilde{a}\dot{\tilde{\varphi}}^2 \big(\frac{\delta\tilde{\varphi}}{\dot{\tilde{\varphi}}}\big),
\eea
and expressing $\frac{\delta\tilde{\varphi}}{\dot{\tilde{\varphi}}}$ and $\tilde{\Psi}$ in terms of $\tilde{\zeta}$ and $\tilde{\xi}$, one obtains
\bea
\dot{\tilde{H}}\tilde{\xi}+\tilde{H}\dot{\tilde{\xi}}=\tilde{a}\dot{\tilde{\varphi}}^2 \Big[\frac{\tilde{\zeta}}{\tilde{H}}-\Big(\frac{1}{2\tilde{a}}-\frac{K}{\tilde{a}^3 \dot{\tilde{\varphi}}^2}\Big)\tilde{\xi}\Big]=\tilde{a}\dot{\tilde{\varphi}}^2 \frac{\tilde{\zeta}}{\tilde{H}}-\Big(\frac{\dot{\tilde{\varphi}}^2}{2}-\frac{K}{\tilde{a}^2}\Big)\tilde{\xi}.
\eea
Then, using the equation $\dot{\tilde{H}}-\frac{K}{\tilde{a}^2}=-\frac{1}{2}\dot{\tilde{\varphi}}^2$ obtained from the background equation~(\ref{bge1}) and ~(\ref{bge2}), one gets
\bea
\dot{\tilde{\xi}}=\tilde{a}\frac{\dot{\tilde{\varphi}}^2}{\tilde{H}^2}\tilde{\zeta},
\eea
Equation~(\ref{ee1}) can be written as
\bea
&&\frac{1}{\tilde{a}^2}(\Delta+3K)\tilde{\Psi}-3\tilde{H}(\dot{\tilde{\Psi}}+\tilde{H}\tilde{\Phi})=\frac{1}{2}\Big[\dot{\tilde{\varphi}}\dot{\delta\tilde{\varphi}}-\tilde{\Phi}\dot{\tilde{\varphi}}^2-\Big(\dot{\tilde{\varphi}}\ddot{\tilde{\varphi}}+3\tilde{H}\dot{\tilde{\varphi}}^2\Big)\frac{1}{\dot{\tilde{\varphi}}}\delta\tilde{\varphi}\Big] \\ \nonumber
&&=\frac{1}{2}\Big[-\dot{\tilde{\varphi}}^2\Big(\tilde{\Phi}-\dot{\tilde{\varphi}}\frac{\dot{\delta\tilde{\varphi}}}{\dot{\tilde{\varphi}}^2}+\ddot{\tilde{\varphi}}\frac{\delta\tilde{\varphi}}{\dot{\tilde{\varphi}}^2}\Big)-3\tilde{H}\dot{\tilde{\varphi}}^2\frac{\delta\tilde{\varphi}}{\dot{\tilde{\varphi}}}\Big] \\ \nonumber
&&=-\frac{1}{2}\Big\{\dot{\tilde{\varphi}}^2\Big[\tilde{\Phi}-\Big(\frac{\delta\tilde{\varphi}}{\dot{\tilde{\varphi}}}\Big)^{.}\Big]+3\tilde{H}\dot{\tilde{\varphi}}^2\Big(\frac{\delta\tilde{\varphi}}{\dot{\tilde{\varphi}}}\Big)\Big\},
\eea
\bea
&&\frac{1}{\tilde{a}^2}(\Delta+3K)\tilde{\Psi}=\frac{1}{2}\dot{\tilde{\varphi}}^2\Big[\Big(\frac{\delta\tilde{\varphi}}{\dot{\tilde{\varphi}}}\Big)^{.}-\tilde{\Psi}\Big]-\frac{3}{2}\tilde{H}\dot{\tilde{\varphi}}^2\big(\frac{\delta\tilde{\varphi}}{\dot{\tilde{\varphi}}}\Big)+3\tilde{H}(\dot{\tilde{\Psi}}+\tilde{H}\tilde{\Psi}) \\ \nonumber
&&=\frac{1}{2}\dot{\tilde{\varphi}}^2\Big[\Big(\frac{\delta\tilde{\varphi}}{\dot{\tilde{\varphi}}}\Big)^{.}-\tilde{\Psi}\Big]-\frac{3}{2}\tilde{H}\dot{\tilde{\varphi}}^2\big(\frac{\delta\tilde{\varphi}}{\dot{\tilde{\varphi}}}\Big)+3\tilde{H}\Big[\frac{1}{2}\dot{\tilde{\varphi}}^2 \Big(\frac{\delta\tilde{\varphi}}{\dot{\tilde{\varphi}}}\Big)\Big] \\ \nonumber
&&=\frac{1}{2}\dot{\tilde{\varphi}}^2\Big[\Big(\frac{\delta\tilde{\varphi}}{\dot{\tilde{\varphi}}}\Big)^{.}-\tilde{\Psi}\Big],
\eea
which gives
\bea
\Big(\frac{\delta\tilde{\varphi}}{\dot{\tilde{\varphi}}}\Big)^{.}=\Big[\frac{2}{\tilde{a}^2 \dot{\tilde{\varphi}}^2}(\Delta+3K)+1\Big]\tilde{\Psi}.
\eea
Expressing $\frac{\delta\tilde{\varphi}}{\dot{\tilde{\varphi}}}$ and $\tilde{\Psi}$ in terms of $\tilde{\zeta}$ and $\tilde{\xi}$, one obtains
\bea
\Big[\frac{\tilde{\zeta}}{\tilde{H}}-\Big(\frac{1}{2\tilde{a}}-\frac{K}{\tilde{a}^3 \dot{\tilde{\varphi}}^2}\Big)\tilde{\xi}\Big]^{.}=\Big[\frac{2}{\tilde{a}^2 \dot{\tilde{\varphi}}^2}(\Delta+3K)+1\Big]\tilde{\Psi},
\eea
\bea
\frac{\dot{\tilde{\zeta}}\tilde{H}-\tilde{\zeta}\dot{\tilde{H}}}{\tilde{H}^2}+\frac{1}{2}\frac{\tilde{H}}{\tilde{a}}\tilde{\xi}+K\Big(\frac{3\tilde{H}}{\dot{\tilde{\varphi}}^2 \tilde{a}^3}+2\frac{\tilde{V}_{\tilde{\varphi}}}{\dot{\tilde{\varphi}}^3 \tilde{a}^3}\Big)\tilde{\xi}-\Big(\frac{1}{2\tilde{a}}-\frac{K}{\dot{\tilde{\varphi}}^3 \tilde{a}^2}\Big)\dot{\tilde{\xi}}=\Big[\frac{2}{\tilde{a}^2 \dot{\tilde{\varphi}}^2}(\Delta+3K)+1\Big]\tilde{\Psi},
\eea
\bea
\frac{\dot{\tilde{\zeta}}}{\tilde{H}}-\dot{\tilde{H}}\frac{\tilde{\zeta}}{\tilde{H}^2}-\Big(\frac{1}{2}\dot{\tilde{\varphi}}^2-\frac{K}{\tilde{a}^2}\Big)\frac{\tilde{\zeta}}{\tilde{H}^2}+\frac{1}{2}\frac{\tilde{H}}{\tilde{a}}\tilde{\xi}+K\Big(\frac{3\tilde{H}}{\dot{\tilde{\varphi}}^2 \tilde{a}^3}+2\frac{\tilde{V}_{\tilde{\varphi}}}{\dot{\tilde{\varphi}}^3 \tilde{a}^3}\Big)\tilde{\xi}=\Big[\frac{\tilde{H}}{\tilde{a}^3 \dot{\tilde{\varphi}}^2}(\Delta+3K)+\frac{\tilde{H}}{2\tilde{a}}\Big]\tilde{\xi},
\eea
which reduces to
\bea
\frac{\dot{\tilde{\zeta}}}{\tilde{H}}+2\frac{\tilde{V}_{\tilde{\varphi}}}{\dot{\tilde{\varphi}}^3 \tilde{a}^3}K\tilde{\xi}=\frac{\tilde{H}}{\tilde{a}^3 \dot{\tilde{\varphi}}^2}\Delta\tilde{\xi},
\eea
and can be rewritten as
\bea
\dot{\tilde{\zeta}}=\frac{H^2}{\tilde{a}^3 \dot{\tilde{\varphi}}^2}\Big(\Delta-2\frac{\tilde{V}_{\tilde{\varphi}}}{\tilde{H}\dot{\tilde{\varphi}}}K\Big)\tilde{\xi}=\frac{\tilde{H}^2}{\tilde{a}^3 \dot{\tilde{\varphi}}^2}(\Delta+YK)\tilde{\xi}.
\eea
where $Y=-2\frac{\tilde{V}_{\tilde{\varphi}}}{\tilde{H}\dot{\tilde{\varphi}}}$.

\begin{acknowledgments}

This work was supported by the National Natural Science Foundation of China under Grants Nos. 11865018, 12265019, the Natural Science Research Project of Education Department of Anhui Province of China under Grants No.2022AH051634, the Doctoral Foundation of Zunyi Normal University of China under Grants No. BS[2017]07.

\end{acknowledgments}

\end{document}